\documentclass[conference]{IEEEtran}
\IEEEoverridecommandlockouts
\usepackage{cite}
\usepackage{amsmath,amssymb,amsfonts}
\usepackage{algorithmic}
\usepackage{graphicx}
\usepackage{textcomp}
\usepackage{graphicx} 
\usepackage{svg}
\usepackage{subcaption} 
\usepackage{tabularx}
\usepackage{xcolor}
\usepackage{authblk}
 \def\BibTeX{{\rm B\kern-.05em{\sc i\kern-.025em b}\kern-.08em
    T\kern-.1667em\lower.7ex\hbox{E}\kern-.125emX}}
\begin{document}

\title{Channel-Adaptive Wireless Image Semantic Transmission with Learnable Prompts}

\author[1]{Liang~Zhang}
\author[1*]{Danlan~Huang\thanks{*This is corresponding author. huangdl@bupt.edu.cn.}}
\author[1]{Xinyi~Zhou}
\author[2]{Feng~Ding}
\author[1]{Sheng~Wu}
\author[1]{Zhiqing~Wei}

\affil[1]{Beijing University of Posts and Telecommunications, Beijing, China}
\affil[2]{Beijing Aerospace Science and Industry Century Satellite High Technology Co., Ltd, Beijing, China}

\maketitle

\begin{abstract}
Recent developments in Deep learning based Joint Source-Channel Coding (DeepJSCC) have demonstrated impressive capabilities within wireless semantic communications system. However, existing DeepJSCC methodologies exhibit limited generalization ability across varying channel conditions, necessitating the preparation of multiple models. Optimal performance is only attained when the channel status during testing aligns precisely with the training channel status, which is very inconvenient for real-life applications. In this paper, we introduce a novel DeepJSCC framework, termed Prompt JSCC (PJSCC), which incorporates a learnable prompt to implicitly integrate the physical channel state into the transmission system. Specifically, the Channel State Prompt (CSP) module is devised to generate prompts based on diverse SNR and channel distribution models. Through the interaction of latent image features with channel features derived from the CSP module, the DeepJSCC process dynamically adapts to varying channel conditions without necessitating retraining. Comparative analyses against leading DeepJSCC methodologies and traditional separate coding approaches reveal that the proposed PJSCC achieves optimal image reconstruction performance across different SNR settings and various channel models, as assessed by Peak Signal-to-Noise Ratio (PSNR) and Learning-based Perceptual Image Patch Similarity (LPIPS) metrics. Furthermore, in real-world scenarios, PJSCC shows excellent memory efficiency and scalability, rendering it readily deployable on resource-constrained platforms to facilitate semantic communications.
\end{abstract}


\section{INTRODUCTION}
In wireless communications, source coding and channel coding are distinct techniques utilized for data compression in the application layer and enhancing channel reliability in the physical layer, respectively. Rooted in Shannon's well-established Separation Theorem \cite{1950The}, these techniques have traditionally been applied separately to achieve theoretically optimal performance when block lengths are infinite. However, this optimal performance is unattainable in practical scenarios, leading to expectations of a substantial paradigm shift in the next generation of communication systems.

Joint Source-Channel Coding (JSCC) facilitates a transformation from the source signal space to the channel input space—and vice versa at the receiver. Notably, JSCC avoids the ``cliff effect" and is recognized for achieving optimal system-level performance in the coding process.
Furthermore, the encoder and decoder can be parameterized using deep neural networks, introducing the so-called DeepJSCC design \cite{bourtsoulatze2019deep,kurka2020deepjscc,peng2022robust}, which is integral to the emerging field of semantic communications. This field has attracted increasing research attention. In wireless image transmission, semantic communications systems aim to transmit the underlying meaning of the source image through the noisy channel with the highest possible fidelity\cite{wang2022perceptual}.
 \begin{figure*}[t]
    \centering
    \includegraphics[width=0.9\textwidth]{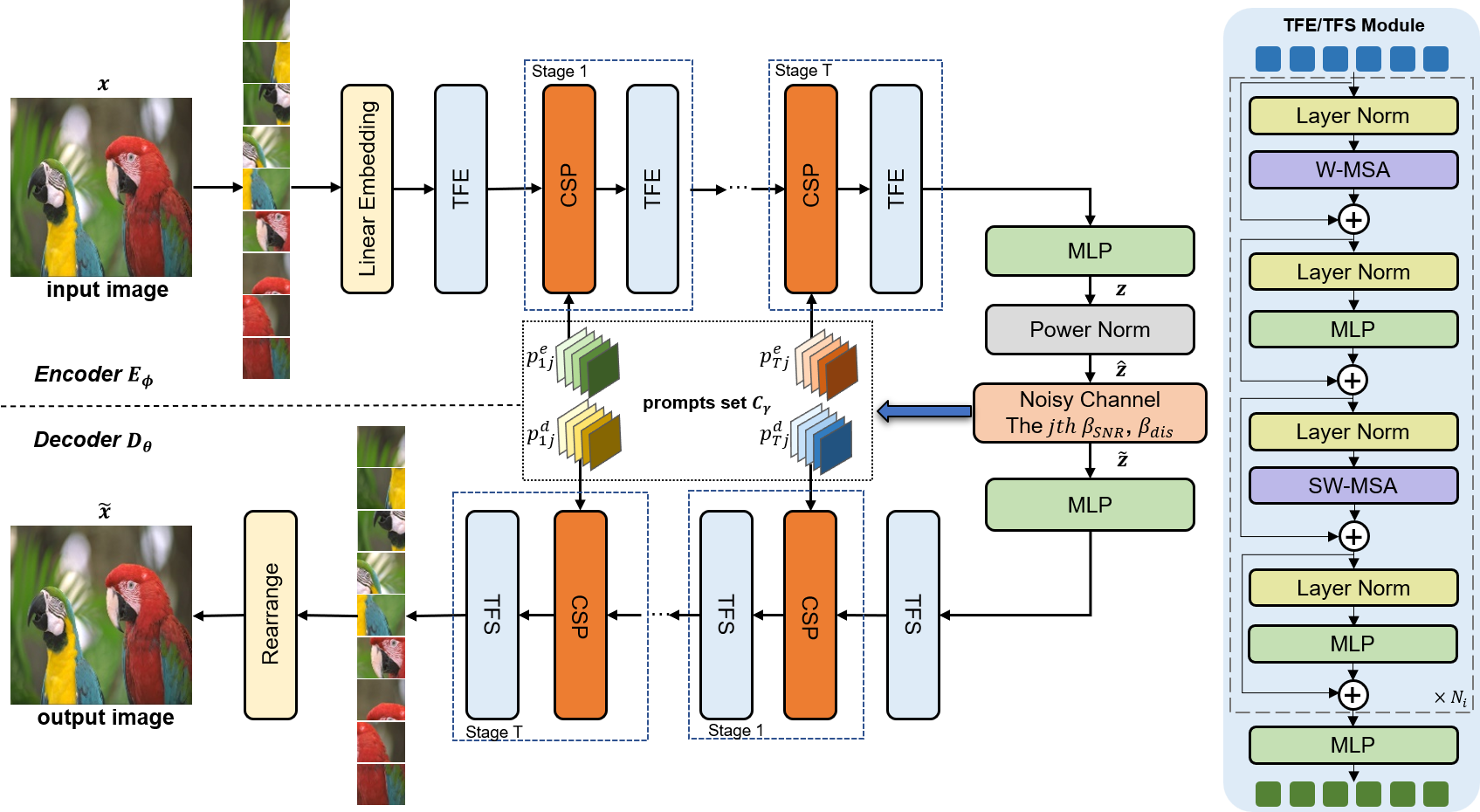}
    \caption{The architecture of the proposed PJSCC for wireless image transmission with channel information prompt.}
    \label{framework}
\end{figure*}

A primary challenge \cite{yang2022semantic} in the design of DeepJSCC lies in its adaptability to varying wireless channel conditions. Typically, existing DeepJSCC models are trained for a specific channel Signal-to-Noise Ratio (SNR), achieving optimal performance primarily when operational channel SNRs closely match the training SNRs. Given that channel SNRs can fluctuate across a broad regimes, from low to high, it is necessary to develop multiple DeepJSCC models to adequately cover the entire SNR range. This requirement significantly escalates training and deployment costs, rendering it impractical for resource-constrained platforms such as mobile and edge devices. Recent works enhance the adaptability of DeepJSCC by directly feeding the channel information to the network, enabling the adjustment of the image's features based on the SNR via fully connected layers \cite{xu2022deep, xu2021wireless, zhang2023predictive}. This architecture is trained over a range of SNR values, and thus in poor channel conditions only the most important features are transmitted with more robustness against channel noise. Although these methods are straightforward to implement and be understood, they exhibit sub-optimal performance due to the static and simplistic nature of SNR prior information. Consequently, existing works are limited in their ability to adapt to different channel conditions.

Transforming various complex channel state information into prior knowledge that is understandable by the model presents a significant challenge. Recent developments in prompting-based methodologies offer a novel approach to provide in-context information for model fine-tuning on specific tasks. Rather than relying on predefined manual instruction sets as prompts, learnable prompts facilitate more parameter-efficient adaptation of models. Prior research has demonstrated that visual and language prompt tuning presents an effective alternative to comprehensive fine-tuning for Transformer models, enabling adaptability to diverse visual contexts and task-specific directions \cite{jia2022visual, lu2024prompt, zhou2022learning}. Inspired from these advances, to tackle the varying channel conditions issue in practical deployment, it is promising to design a learnable wireless channel prompt to efficiently tuning the DeepJSCC model from the physical layer aspect. By introducing a modest number of trainable parameters within the input space, a universal model for wireless image transmission can be devised, capable of dynamically recovering inputs amidst varying levels of channel noise through their interaction with the prompts. These prompts function as adaptive, lightweight modules tasked with SNR levels and channel models within the DeepJSCC network.

This paper presents Prompt Joint Source-Channel Coding (PJSCC), an extension of the DeepJSCC model, introducing learnable prompts to achieve channel-adaptive capability in wireless image semantic transmission. The primary contributions are summarized as follows.
\begin{itemize}
\item A Channel State Prompting (CSP) module is proposed to generate prompts customized to the channel conditions, considering SNR and channel distribution model. These prompts are intricately fused with image features extracted from preceding encoders. Consequently, the fluctuating channel environment interacts with the transmitted image within the network architecture. 
\item The proposed PJSCC model demonstrates remarkable scalability in accommodating diverse noise levels in both Additive White Gaussian Noise (AWGN) and Rayleigh fading channel models. This advancement alleviates the burden of model retraining and deployment, particularly in resource-constrained platforms.
\item  Extensive experiments demonstrate the significant improvement in image reconstruction of the proposed PJSCC, in terms of Peak Signal-to-Noise Ratio (PSNR) and Learned Perceptual Image Patch Similarity (LPIPS).
\end{itemize}


\section{The Proposed Channel-Adaptive PJSCC Model}
In this section, we elaborate the system model, the proposed encoder, decoder, CSP module, and model training.
\subsection{System Model}
We investigate a wireless image semantic transmission system leveraging a DeepJSCC scheme combined with a Transformer architecture, as detailed in\cite{dosovitskiy2020image}. This system demonstrates exceptional ability to model extensive dependencies and exhibits scalability in processing vision information. The primary objective is the accurate reconstruction of the original input image at the decoder end, from the received semantic code (SC), under specific SNR and Channel Bandwidth Ratio (CBR) conditions. The overall system model of the proposed PJSCC is illustrated in Fig. \ref{framework}, which is composed of a trainable encoder $\pmb{E}_{\phi}$, a trainable decoder $\pmb{D}_\theta$, and a set of trainable prompt parameters $\pmb{C}_\gamma$ that introduces physical channel parameters into the transmission system. Let the input image to be transmitted be denoted as $\pmb{x}\in \mathbb{R}^{3\times H \times W}$, where 3 represents the three RGB channels, and $H, W$ denotes the height and width of the image, respectively. The input image $\pmb{x}$ will be firstly encoded as a $K$-dimensional complex SC $\pmb{z}\in \mathbb{C}^K$ as follows:
\begin{equation}
\pmb{z}=\pmb{E}_\phi(\pmb{x}, \pmb{C}_\gamma(\beta_{SNR}, \beta_{dis}), R),
\end{equation}
where $\beta_{SNR}$ means channel SNR, and $\beta_{dis}$ means the channel distribution type, such as AWGN or Rayleigh fading distribution. Additionally, $\pmb{C}_\gamma$ is the learned channel state prompts set, and $R=K/(3\times H \times W)$ denotes the CBR condition. 

Subsequently, the SC $\pmb{z}$ undergoes power normalization before being transmitted through the physical channel, as given by:
\begin{equation}
\pmb{\hat{z}} = \sqrt{KP}\frac{\pmb{z}}{\sqrt{\pmb{z}^*\pmb{z}}},
\end{equation}
where $\pmb{z}^*$ represents the conjugate transpose of $\pmb{z}$, and $P$ denotes the average transmission power constraint. Subsequently, the normalized SC $\pmb{\hat{z}}$ is projected as a complex symbol. 
We consider both the AWGN and the Rayleigh fading channel models. The AWGN transmitting model is given by:
\begin{equation}
\pmb{\tilde{z}} = \pmb{\hat{z}}+\pmb{n},
\end{equation}
where $\pmb{n}\sim\mathcal{CN}(0,\sigma^2\pmb{I}_K)$ denotes independent identically distributed (i.i.d) samples from a circularly symmetric complex Gaussian distribution with noise power $\sigma^2$. Additionally, the Rayleigh fading channel model is given as:
\begin{equation}
\pmb{\tilde{z}} = \pmb{h}\pmb{\hat{z}}+\pmb{n},
\end{equation}
where the channel gain is $\pmb{h}\sim \mathcal{CN}(0, I_K)$.
The corrupted SC $\pmb{\tilde{z}}$ is processed by the decoder $\pmb{D}_\theta$ at the receiver. The recovered image is given by:
\begin{equation}
\pmb{\tilde{x}}=\pmb{D}_\theta(\pmb{\tilde{z}}, \pmb{C}_\gamma(\beta_{SNR},\beta_{dis}),R).
\end{equation}

The PJSCC model is trained in an end-to-end manner. The loss function is defined as the Mean Squared Error (MSE) distortion between $\pmb{x}$ and $\pmb{\tilde{x}}$, which is formulated as:
\begin{equation}
\mathcal{L}_{\phi,\theta,\gamma}(\pmb{x},\pmb{\tilde{x}}; \pmb{C}_\gamma(\beta_{SNR},\beta_{dis}), R) = ||\pmb{x}-\pmb{\tilde{x}}||^2_2.
\end{equation}
The optimal model parameters can be obtained by:
\begin{equation}
\begin{split}
&\quad(\phi^*,\theta^*,\gamma^*) \\
&\!=\!\mathop{\arg\min}\limits_{\phi,\theta,\gamma}\mathbb{E}_{\pmb{x},\pmb{C}_\gamma, R}[\mathcal{L}_{\phi,\theta,\gamma}(\pmb{x},\pmb{\tilde{x}};\pmb{C}_\gamma(\beta_{SNR},\beta_{dis}), R)],
\end{split}
\end{equation}
where $\mathbb{E}$ denotes the expectation, and $\theta^*,\phi^*$ and $\gamma^*$ represent the optimal parameters for the encoder, decoder, and prompts, respectively. 

We adopt the PSNR and LPIPS as the objective metrics to evaluate the reconstruction performance. For ease of representation, we apply logarithmic transformation to the LPIPS value, as given by:
\begin{equation}
    LogLPIPS(\pmb{x},\pmb{\tilde{x}}) = 10\cdot log_{10} LPIPS(\pmb{x},\pmb{\tilde{x}}).
\end{equation}

\subsection{The Proposed Encoder, Decoder, and CSP Modules}
Existing channel adaptation methods simply concatenate SNR to features and interact with the SNR information through a fully connected layer, enabling the model to obtain channel information. However, this design is overly simplistic, and the distribution knowledge obtained from the input features is insufficient. Additionally, this design lacks scalability and becomes challenging to manually design reasonable inputs when facing more complex channel states.

The objective of the proposed PJSCC system is to establish a unified model capable of effectively handling diverse channel conditions, including both AWGN and Rayleigh fading channels, without the necessity for retraining. We incorporate the Swin Transformer \cite{liu2021swin} architecture for feature extraction at the transmitter and feature synthesis at the receiver. The encoder $\pmb{E}_\phi$ is composed of Transformer Feature Extraction (TFE) and CSP modules, while the decoder $\pmb{D}_\theta$ is composed of Transformer Feature Synthesis (TFS) and CSP modules.

\begin{figure}[t]
  \centering
  \includegraphics[width=0.5\textwidth]{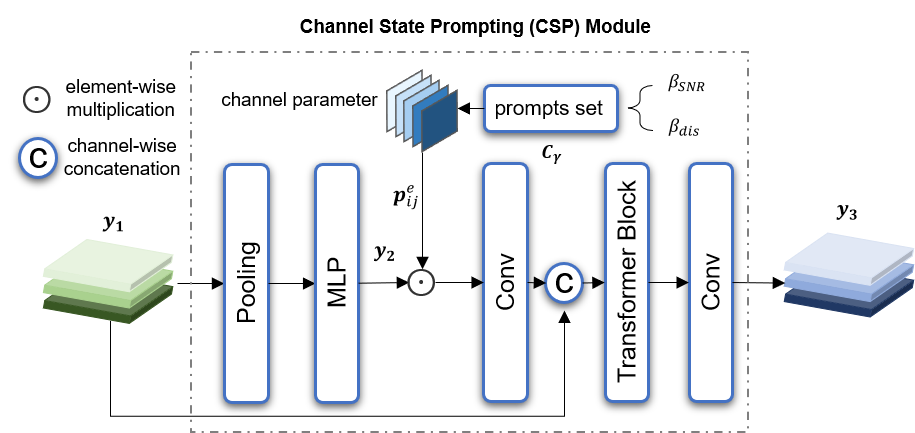}
  \caption{The architecture of CSP module.}
  \label{fig:2}
\end{figure}

The input image $\pmb{x}$ is first partitioned into $M$ non-overlapping patches $\pmb{x}_\textit{p}^\textit{1}, \pmb{x}_\textit{p}^\textit{2}, ..., \pmb{x}_\textit{p}^\textit{M}$ and then being linear embeded into the feature space by matrix $\pmb{E}$ and added up with position embedding $\pmb{E}_\textit{pos}$. 
In our approach, we define two successive layers of the Swin Transformer as a single block, and the subsequent $i$-th TFE module consists of $N_i$ layers of Swin Transformer blocks followed by an MLP layer, which acts as a downsampling layer to reduce the feature dimensions. The output of the TEF module can be formulated as:
\begin{equation}
\pmb{y}_1 = \rm{TFE}([\pmb{x}_\textit{p}^\textit{1}\pmb{E}, \pmb{x}_\textit{p}^\textit{2}\pmb{E},..., \pmb{x}_\textit{p}^\textit{M} \pmb{E}]+\pmb{E}_\textit{pos}).
\end{equation}

The architecture of the proposed CSP module is illustrated in Fig. \ref{fig:2}, with the encoder serving as an example. The CSP module is designed to inject channel information into the image feature extracted from the previous encoder layers, Firstly, in order to lightweight the entire module, we will incorporate features through a Global Average Pooling (GAP) layer and an MLP layer for reshaping. The output latent representation of the MLP layer can be formulated as:
\begin{equation}
\pmb{y}_2=\rm{MLP}\{GAP(\pmb{y}_1)\}.
\end{equation}

Instead of using fixed SNR values directly as input for PJSCC like existing schemes \cite{xu2022deep, xu2021wireless, zhang2023predictive, yang2023witt}, we novelty employ learnable prompts for more parameter-efficient model adaptation.

\begin{figure*}[t]
  \centering
  \begin{subfigure}[b]{0.16\linewidth}
    \includegraphics[width=\linewidth]{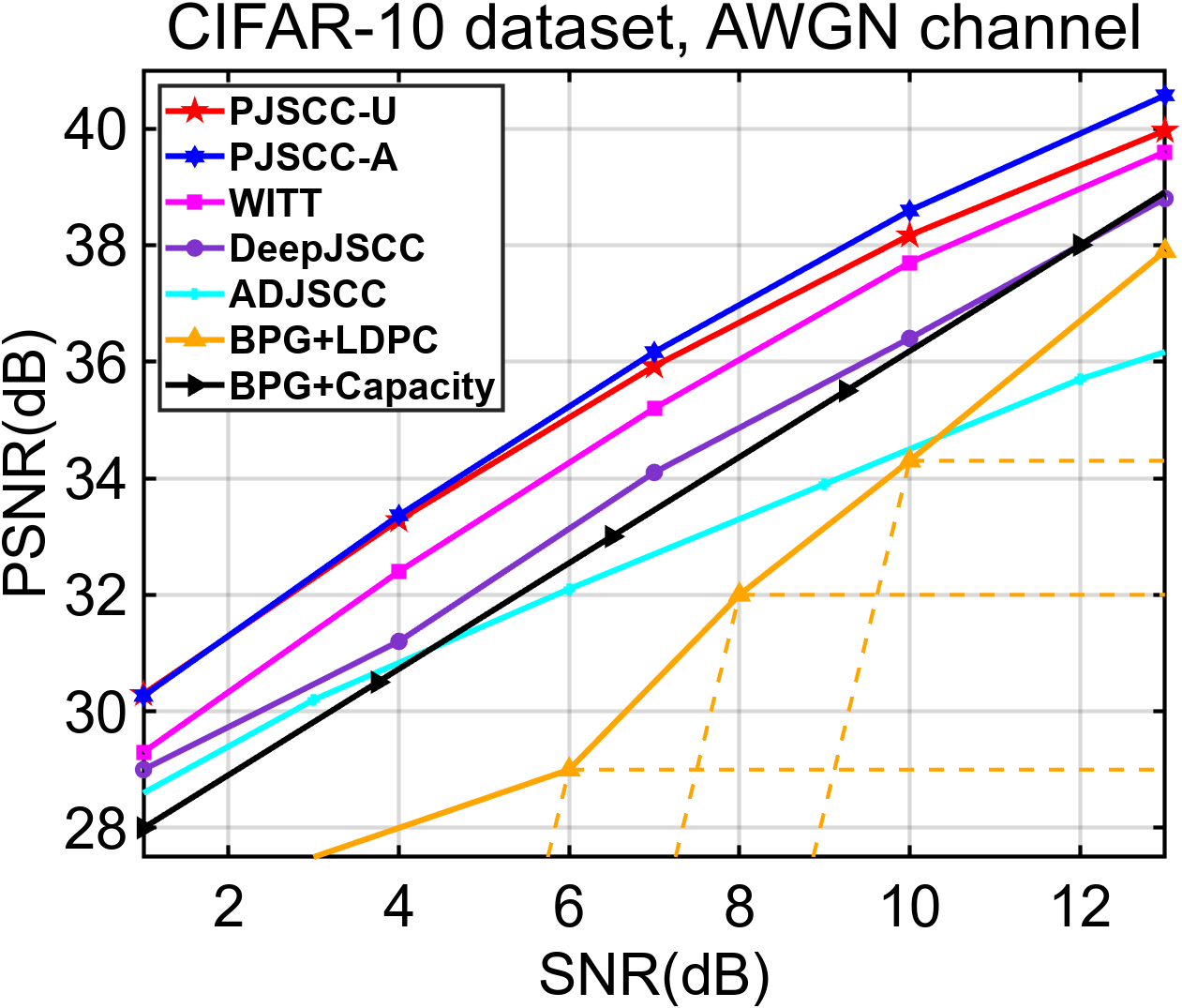}
    \caption{}
  \end{subfigure}
  \hfill
  \begin{subfigure}[b]{0.16\linewidth}
    \includegraphics[width=\linewidth]{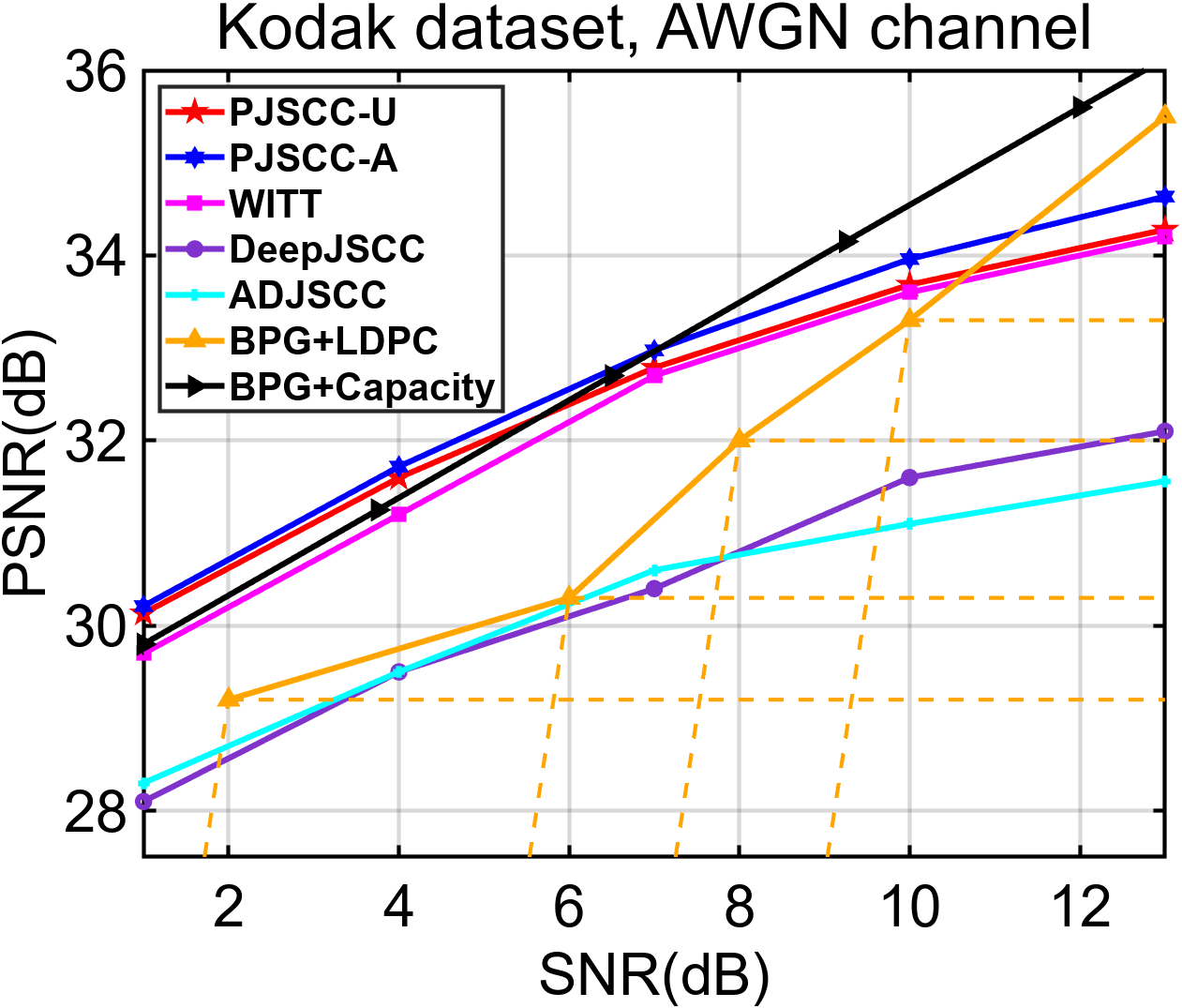}
    \caption{}
  \end{subfigure}
  \hfill
  \begin{subfigure}[b]{0.16\linewidth}
    \includegraphics[width=\linewidth]{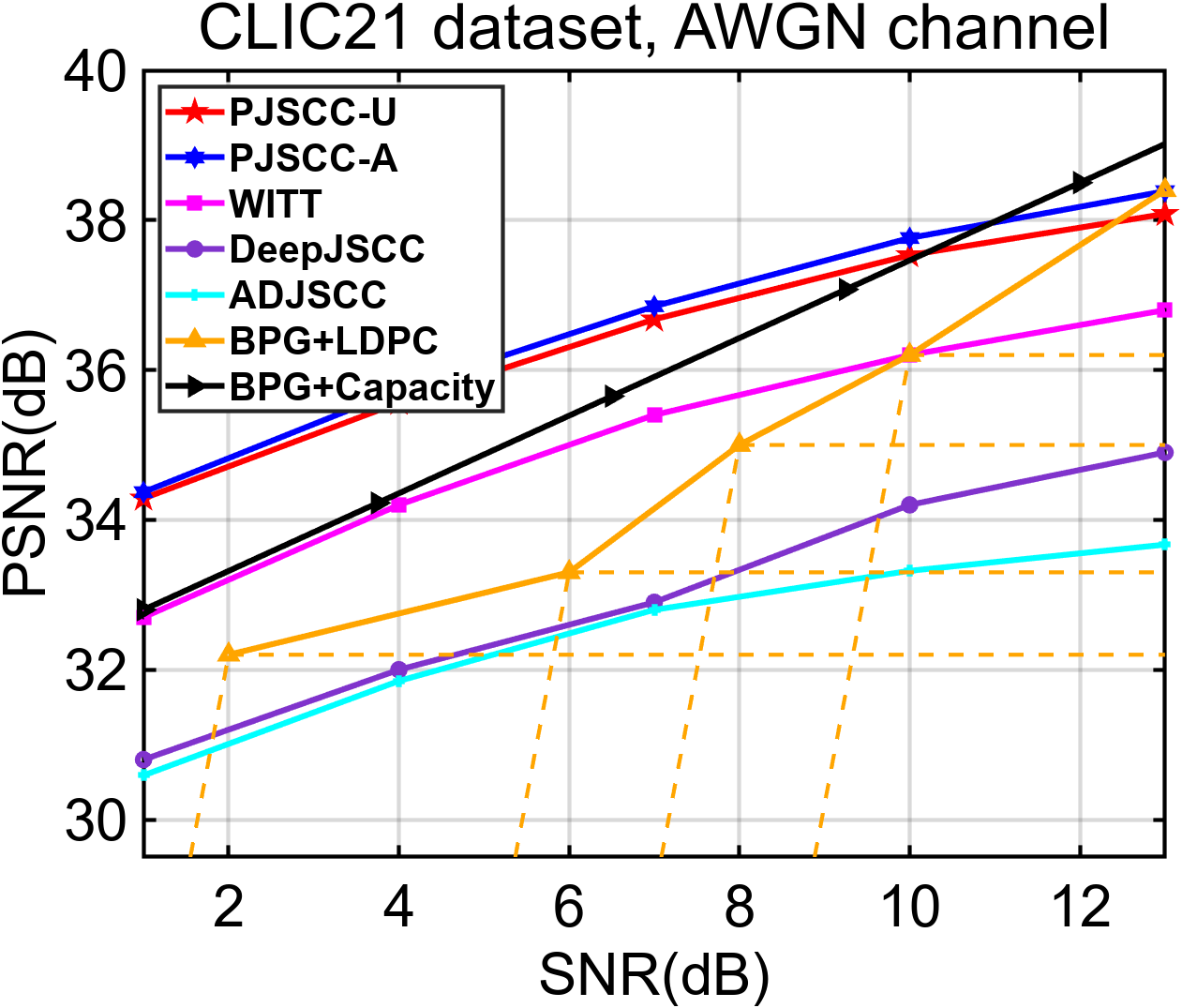}
    \caption{}
  \end{subfigure}
 \hfill
  \begin{subfigure}[b]{0.16\linewidth}
    \includegraphics[width=\linewidth]{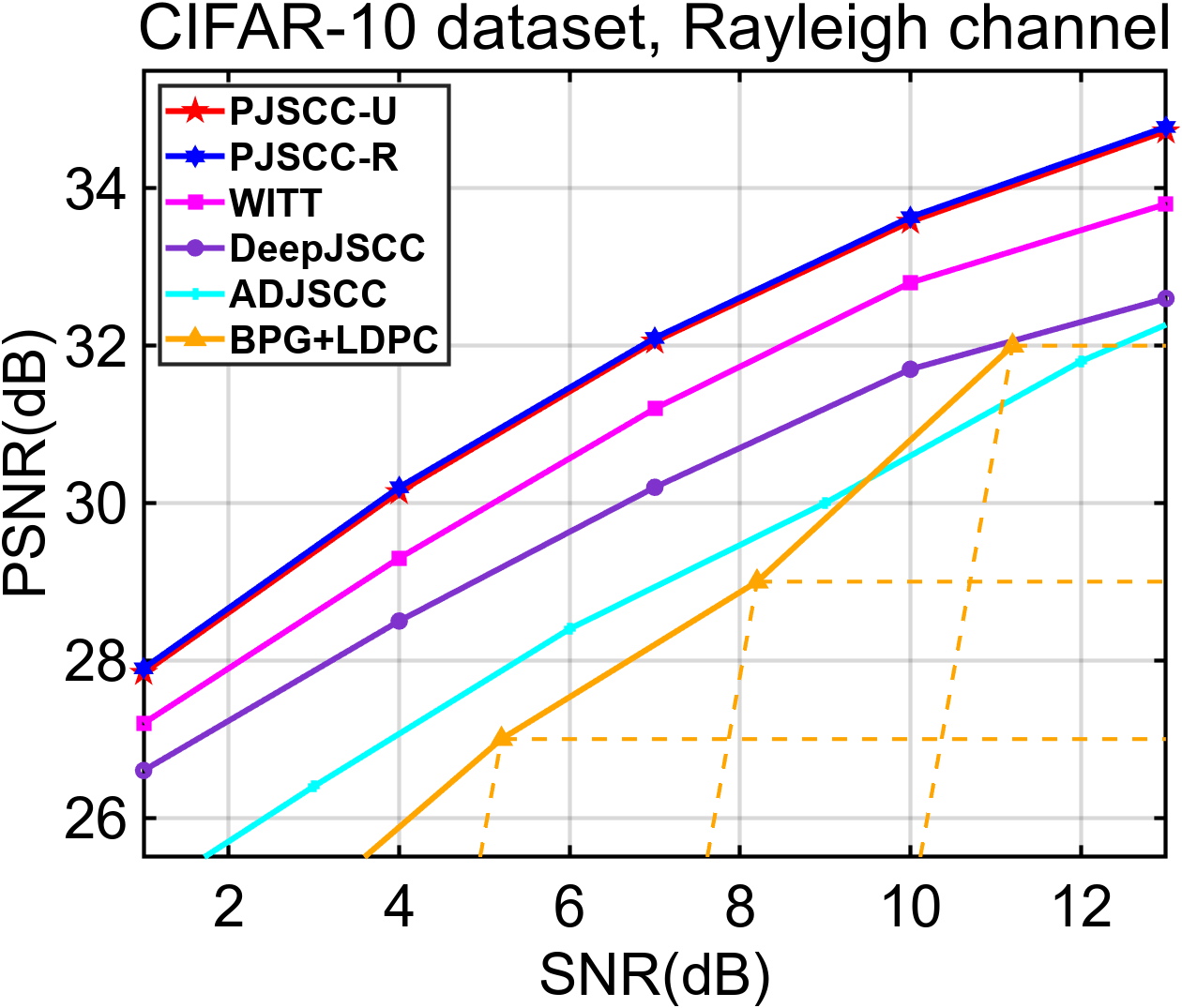}
    \caption{}
  \end{subfigure}
  \hfill
  \begin{subfigure}[b]{0.16\linewidth}
    \includegraphics[width=\linewidth]{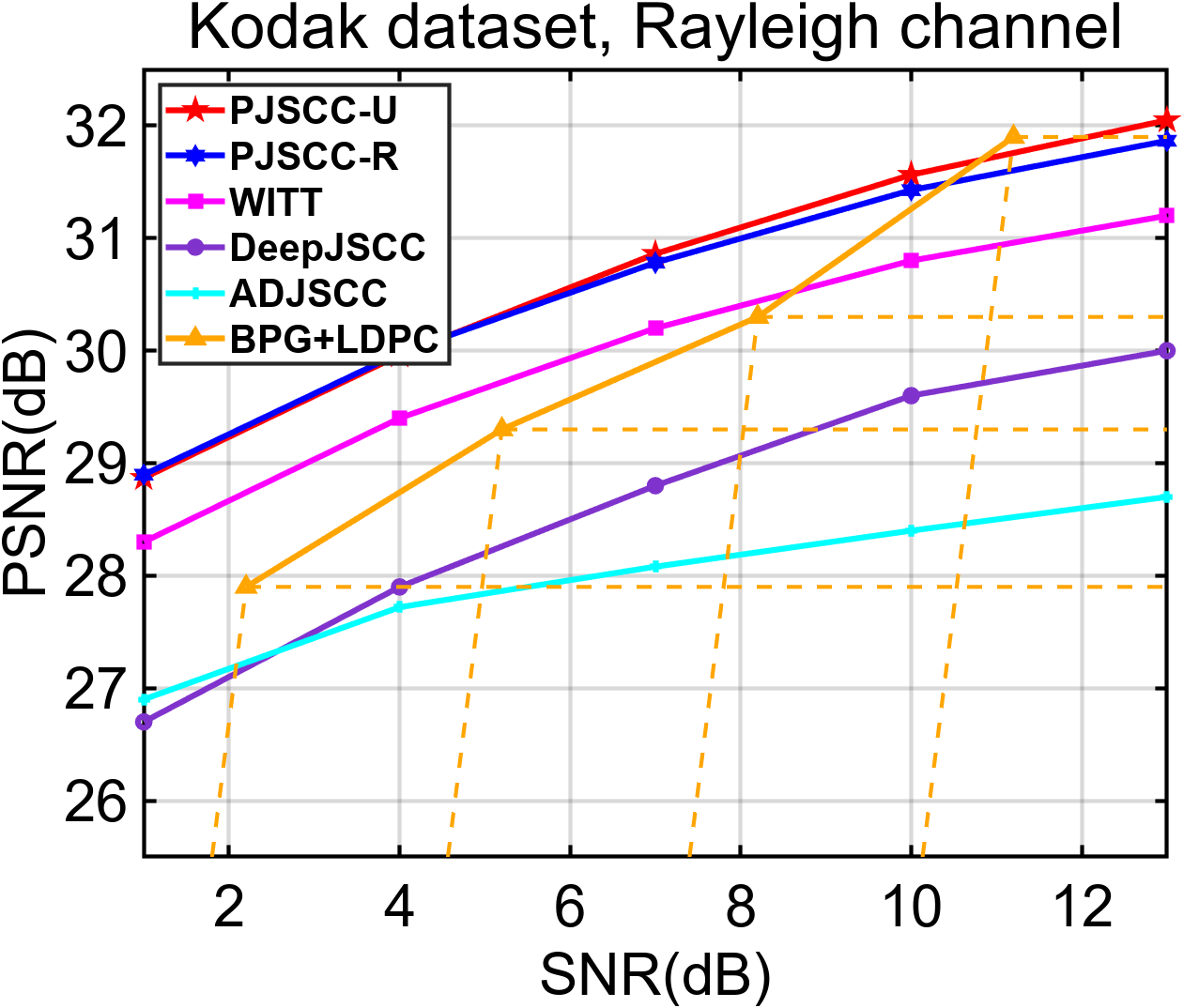}
    \caption{}
  \end{subfigure}
  \hfill
  \begin{subfigure}[b]{0.16\linewidth}
    \includegraphics[width=\linewidth]{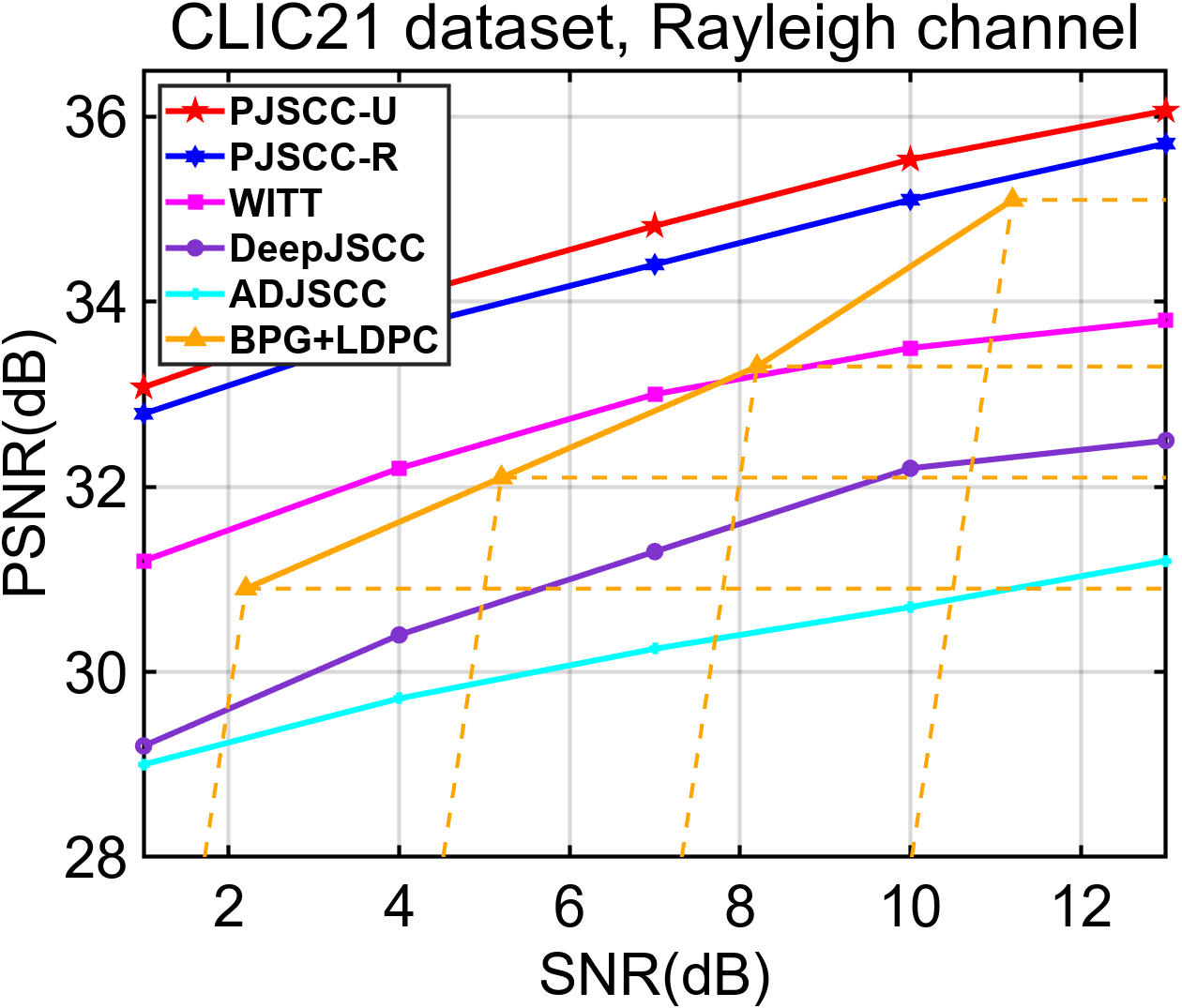}
    \caption{}
  \end{subfigure}
  \caption{The PSNR performance versus various SNRs on CIFAR-10, Kodak, and CLIC datasets, where the CBRs for the three datasets are 1/3 , 1/16 and 1/16, respectively. (a), (b), (c) AWGN channels, (d), (e), (f) Rayleigh fading channels.}
  \label{fig:PSNR}
\end{figure*}
\begin{figure*}[]
  \centering
  \begin{subfigure}[b]{0.16\linewidth}
    \includegraphics[width=\linewidth]{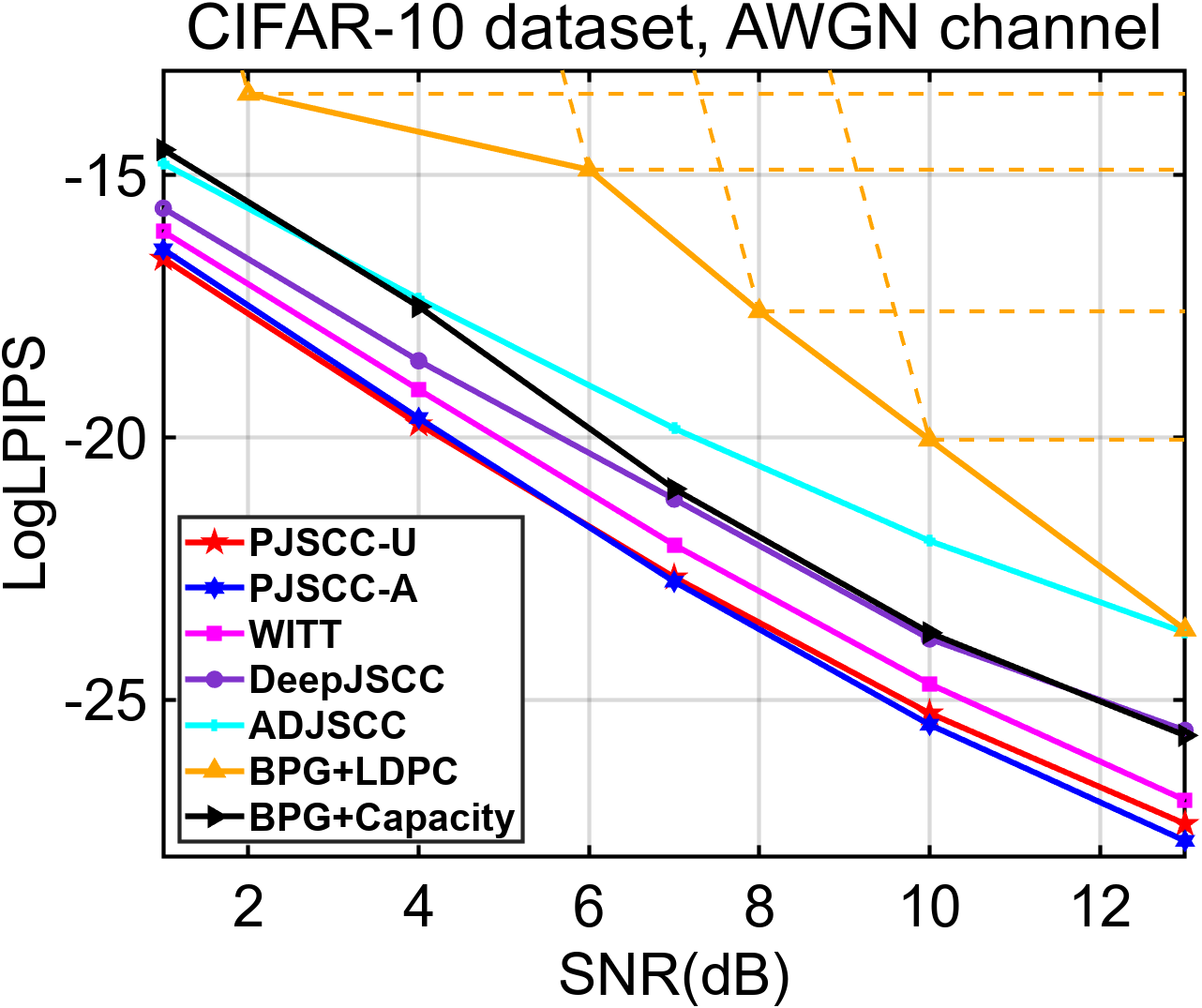}
    \caption{}
  \end{subfigure}
  \hfill
  \begin{subfigure}[b]{0.16\linewidth}
    \includegraphics[width=\linewidth]{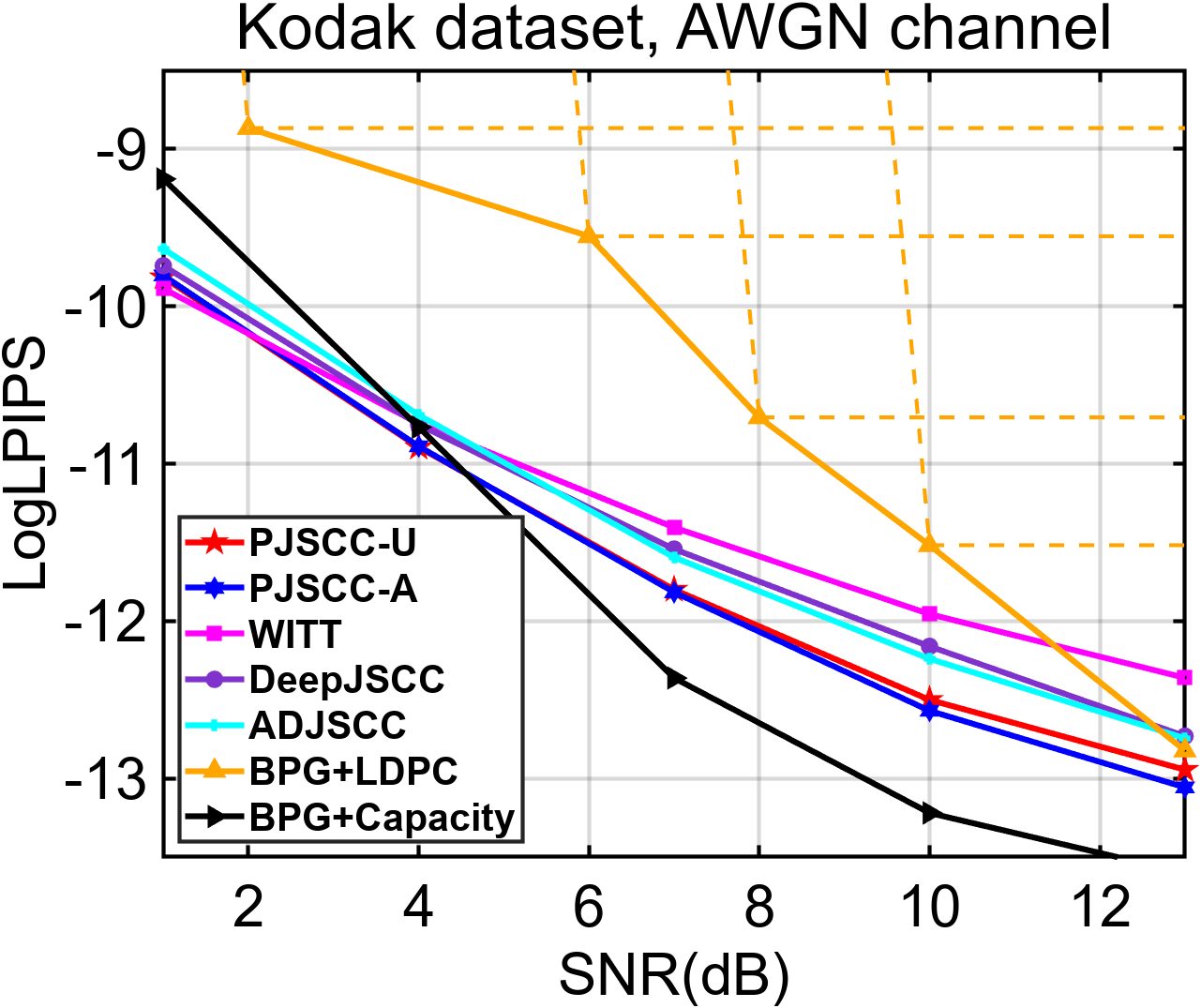}
    \caption{}
  \end{subfigure}
  \hfill
  \begin{subfigure}[b]{0.16\linewidth}
    \includegraphics[width=\linewidth]{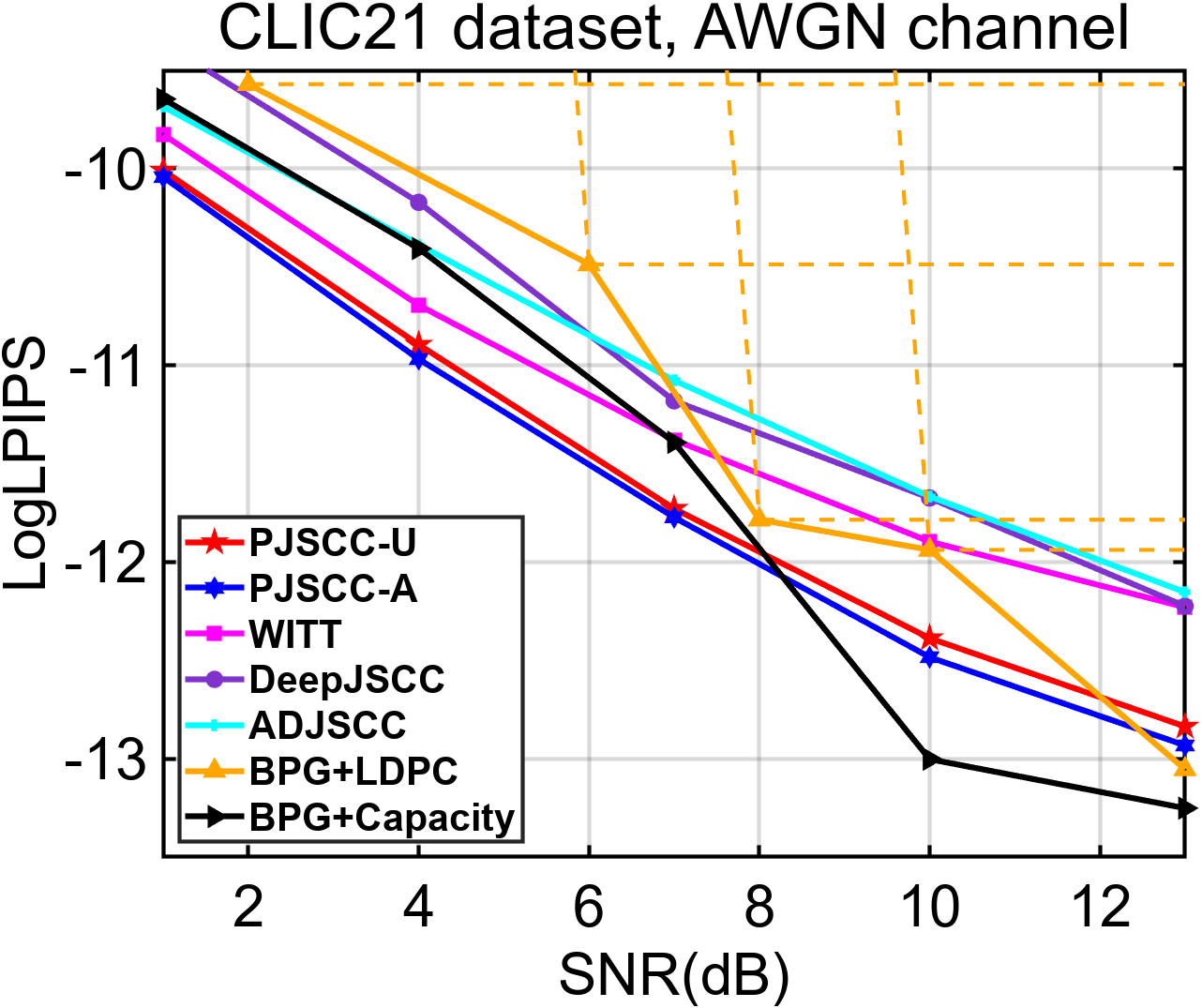}
    \caption{}
  \end{subfigure}
  \hfill
  \begin{subfigure}[b]{0.16\linewidth}
    \includegraphics[width=\linewidth]{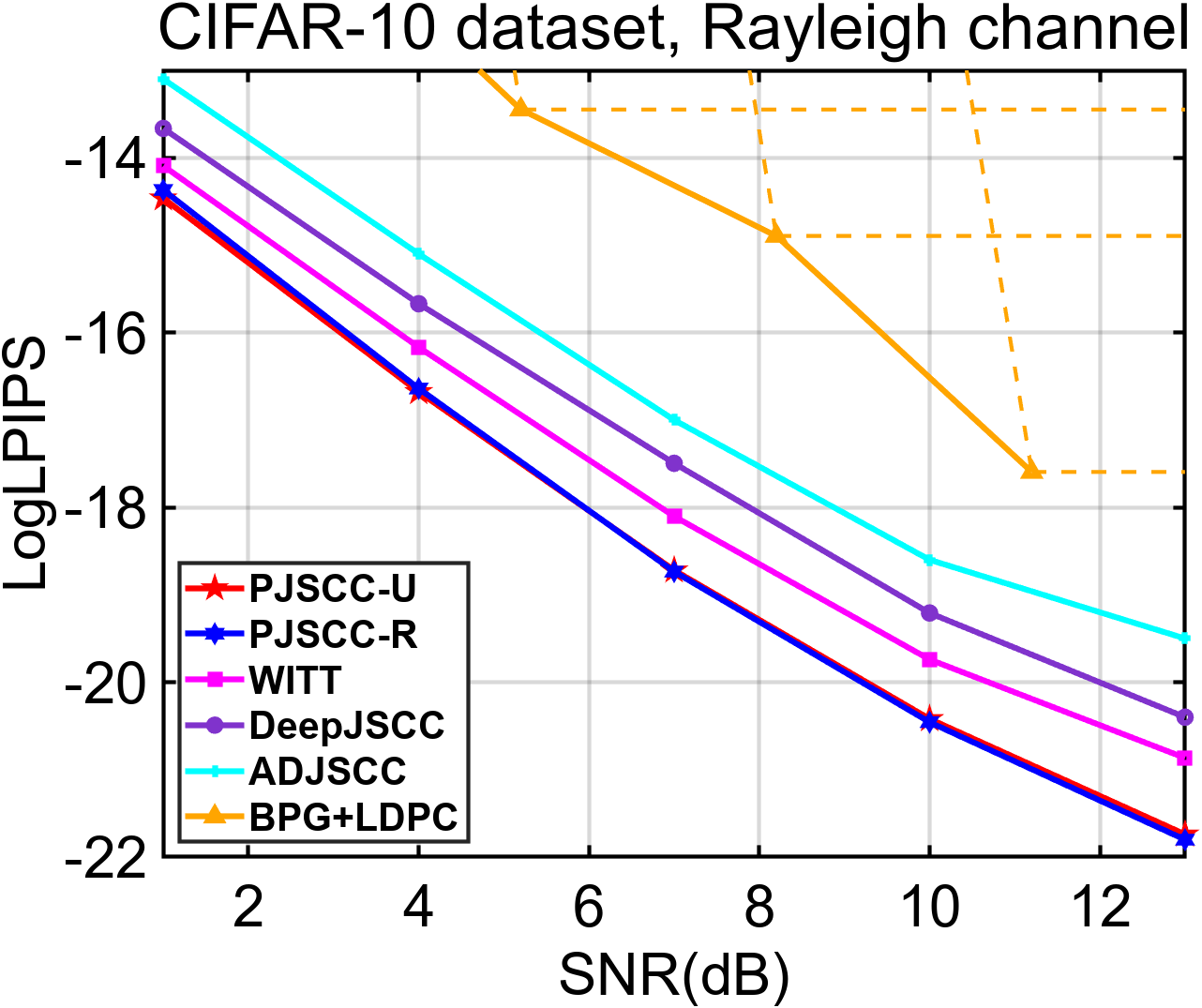}
    \caption{}
  \end{subfigure}
  \hfill
  \begin{subfigure}[b]{0.16\linewidth}
    \includegraphics[width=\linewidth]{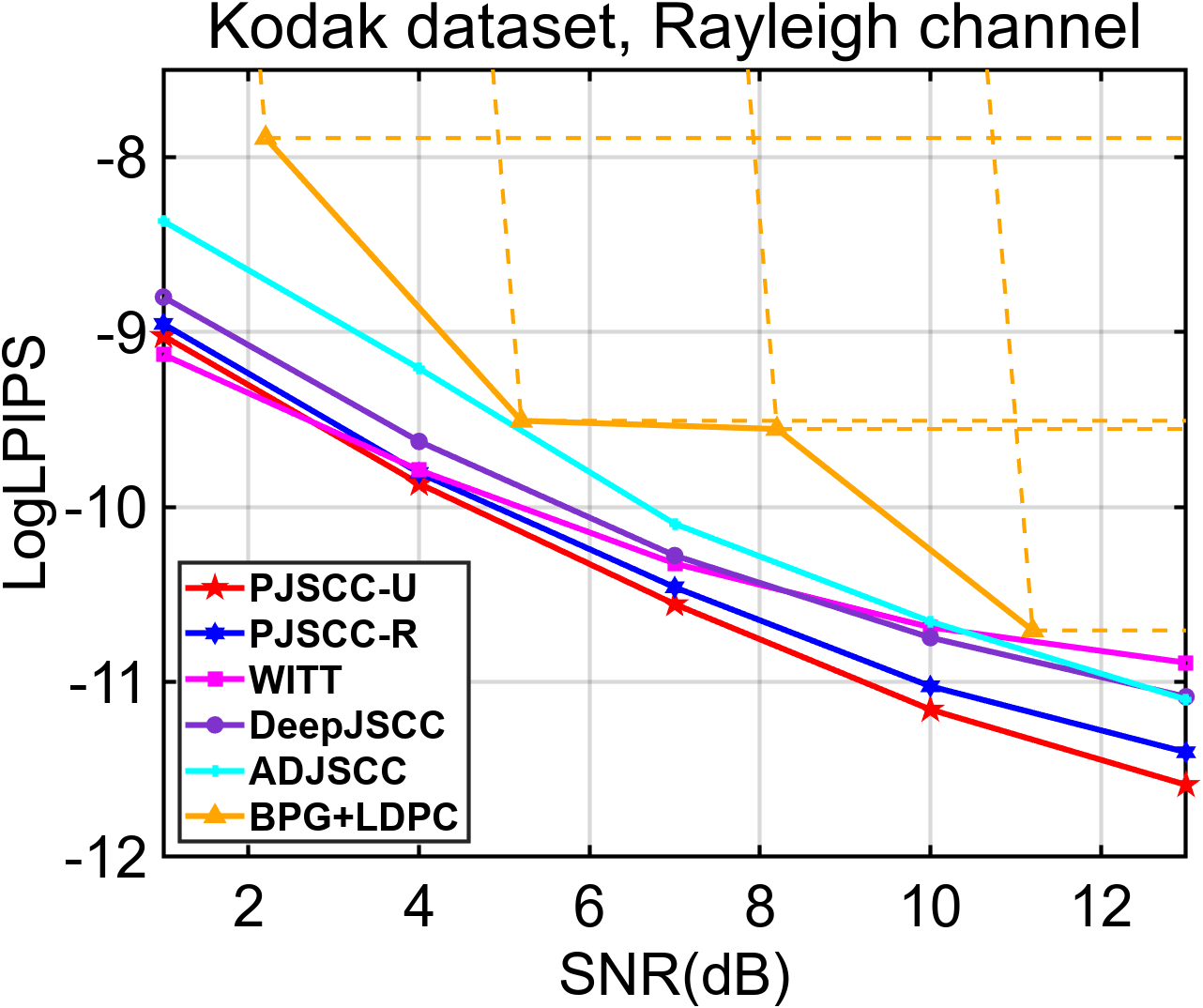}
    \caption{}
  \end{subfigure}
  \hfill
  \begin{subfigure}[b]{0.16\linewidth}
    \includegraphics[width=\linewidth]{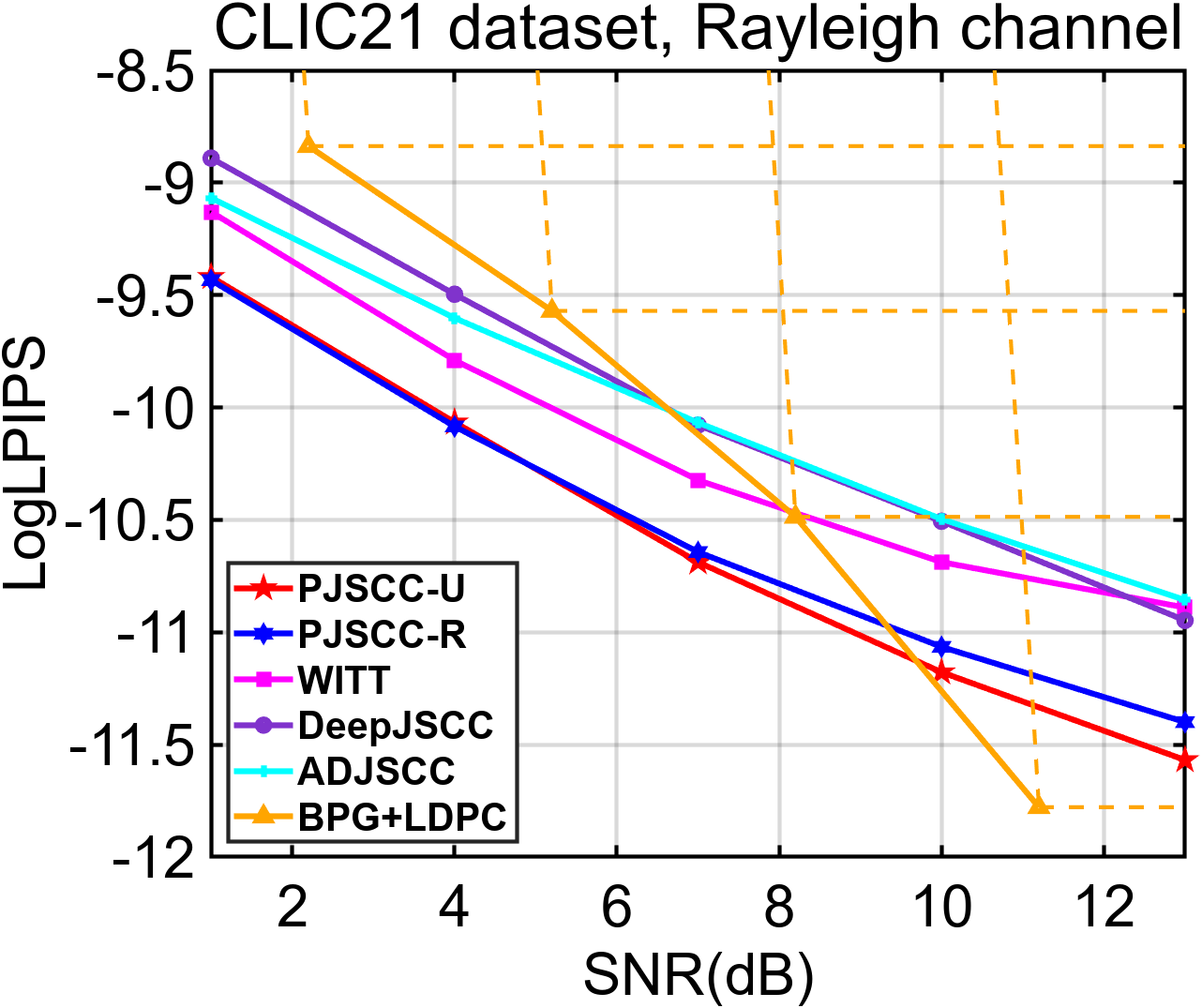}
    \caption{}
  \end{subfigure}
  \caption{The LPIPS performance versus various SNRs on CIFAR-10, Kodak, and CLIC datasets, where the CBRs for the three datasets are 1/3 , 1/16 and 1/16, respectively. (a), (b), (c) AWGN channels, (d), (e), (f) Rayleigh fading channels. }
  \label{fig:LPIPS}
\end{figure*}

The prompts set, denoted as $\pmb{C}_\gamma$, consists of tensors $\{\pmb{p}_{ij}^e,\pmb{p}_{ij}^d\}_{i=1,\ldots,L}^{j=1,\ldots,T}$, where $\pmb{p}_{ij}^e$ and $\pmb{p}_{ij}^d$ are used for the encoder and decoder, respectively. Here, $T$ corresponds to the number of stages, and $L$ to the number of wireless channel states. Initially, $\pmb{p}_{ij}^e\in \mathbb{R}^{c\times h \times w}$ is randomly initialized as a learnable parameter. Each prompt is associated with both the SNR $\beta_{SNR}$ and the channel distribution $\beta_{dis}$, allowing for different prompts under varying channel conditions. Thus, for a particular channel distribution, the prompts are initialized as follows:
\begin{equation}
\{\pmb{p}_{j}^e,\pmb{p}_{j}^d,...,\pmb{p}_{Tj}^e, \pmb{p}_{Tj}^d\}=\pmb{C}_\gamma(\beta_{SNR}, \beta_{dis}).
\end{equation}

The prompt $\pmb{p}_{ij}^e$ interacts with the latent representation $\pmb{y}_2$ by multiplication process followed by a convolutional operation. Through this process, the channel state information is combined with brief feature information to generate specific channel state prompts. 

Next, this channel status prompt is concatenated with the original feature $\pmb{y}_1$ yield by the TFE module, and undergoes one subsequent Transformer block and convolution operation, as given by:
\begin{equation}
\pmb{y}_3\!=\! \rm{Conv}\{TransBlock(\rm{Concat}[\rm{Conv}(\pmb{y}_2\odot \pmb{p}_{\textit{ij}}^\textit{e}),\pmb{y}_1])\},
\end{equation}
where $\odot$ represents element-wise multiplication. Note that the obtained $\pmb{y}_3$ fuses the characteristics of the physical channel information and the image latent representation intricately, and therefore the model dynamically adapts the input image to the physical channel. Next, $\pmb{y}_3$ undergoes another TFE module. A single stage is defined as one CSP module followed by a TFE module. After traversing $T$ such stages, the features are processed by an MLP layer to meet the CBR requirements, and then projected as complex symbols into the physical channel.

The decoder recovers the image from the corrupted SC $\pmb{\tilde{z}}$, and it  exhibits a structure symmetrical to that of the encoder. The main distinction lying in the replacement of TFE module with TFS module, which adopts upsampling layers other than downsampling layers.
The encoder, decoder, and CSP modules are trained in an end-to-end manner, which enables the proposed PJSCC model to dynamically learn the probability distribution of channel states and the input image, as well as optimizing the initial prompts.

\begin{figure*}[t]
  \centering
  \begin{subfigure}[b]{0.19\linewidth}
    \caption*{input image}
    \includegraphics[width=\linewidth]{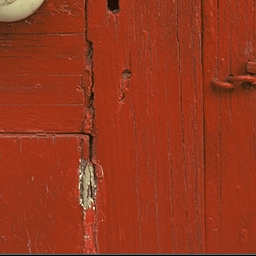}
    \caption*{PSNR/LPIPS}
  \end{subfigure}
  \hfill
  \begin{subfigure}[b]{0.19\linewidth}
    \caption*{PJSCC}
    \includegraphics[width=\linewidth]{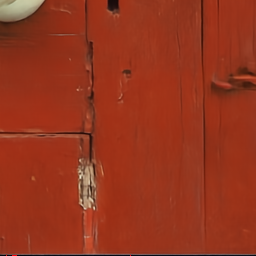}
    \caption*{33.77dB/0.0458}
  \end{subfigure}
  \hfill
  \begin{subfigure}[b]{0.19\linewidth}
  \caption*{WITT}
    \includegraphics[width=\linewidth]{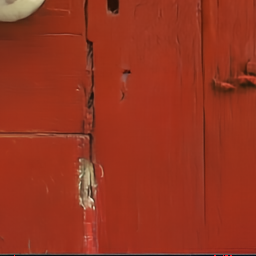}
    \caption*{33.22dB/0.0478}
  \end{subfigure}
  \hfill
  \begin{subfigure}[b]{0.19\linewidth}
  \caption*{ADJSCC}
    \includegraphics[width=\linewidth]{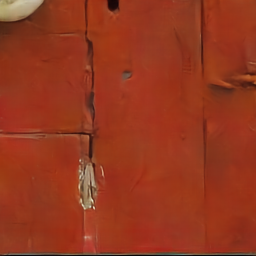}
    \caption*{32.35dB/0.0534}
  \end{subfigure}
  \hfill
  \begin{subfigure}[b]{0.19\linewidth}
  \caption*{BPG+LDPC}
    \includegraphics[width=\linewidth]{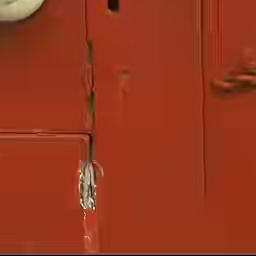}
    \caption*{33.28dB/0.0650}
  \end{subfigure}
  \hfill
  \begin{subfigure}[b]{0.19\linewidth}
    \includegraphics[width=\linewidth]{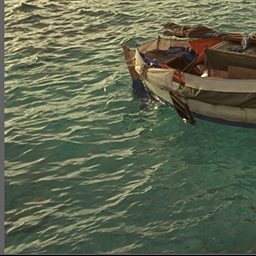}
    \caption*{PSNR/LPIPS}
  \end{subfigure}
  \hfill
  \begin{subfigure}[b]{0.19\linewidth}
    \includegraphics[width=\linewidth]{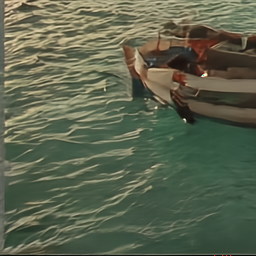}
    \caption*{30.33dB/0.0420}
  \end{subfigure}
  \hfill
  \begin{subfigure}[b]{0.19\linewidth}
    \includegraphics[width=\linewidth]{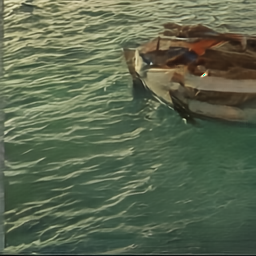}
    \caption*{30.09dB/0.0491}
  \end{subfigure}
  \hfill
  \begin{subfigure}[b]{0.19\linewidth}
    \includegraphics[width=\linewidth]{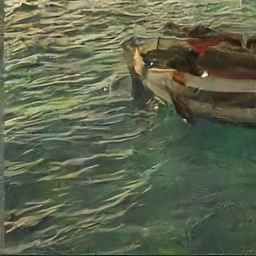}
    \caption*{28.72dB/0.0576}
  \end{subfigure}
  \hfill
  \begin{subfigure}[b]{0.19\linewidth}
    \includegraphics[width=\linewidth]{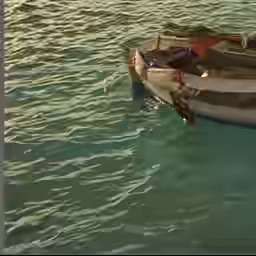}
    \caption*{30.26dB/0.0482}
  \end{subfigure}
  \caption{Visualization of detailed reconstruction results of different semantic coding schemes with CBR = 1/16 and SNR=1dB.}
  \label{fig:Visualization}
\end{figure*}

\section{Experimental Results}
\subsection{Experimental Setup}
\noindent\textbf{1)Datasets}

For low-resolution images, the proposed PJSCC was trained and tested on the CIFAR-10 dataset, which consists of images with a resolution of $32\times32$ pixels. For high-resolution images, we used the DIV2K dataset, which comprises 2K resolution images. During training, we cropped the images into $256\times256$ blocks before feeding them into the network. To verify the generalization capability of our network, we conducted tests on both the Kodak dataset and the CLIC dataset, which include images of varying sizes.

\noindent\textbf{2)Training Details}

As mentioned earlier, in order for the PJSCC model to learn the distribution of different channels, it was trained from scratch under AWGN and Rayleigh fading channel conditions. The channel model was uniformly selected, and the SNR was uniformly distributed between 1 dB and 13 dB. 
In the experimental results, we refer to this universal model as PJSCC-U. We also provide single channel PJSCC models, with PJSCC-A specifically trained for AWGN and PJSCC-R for Rayleigh. For low-resolution images, the level $T$ of the model is set to 1, which includes a CSP module, and the number of Swin Transformer blocks in TFE/TFS is configured as $[N_1, N_2] = [1, 2]$. For high-resolution images, the level $T$ is set to 3, consisting of three CSP modules, with the number of blocks in the TFE/TFS set to $[N_1, N_2, N_3, N_4] = [1, 1, 3, 1]$. For each prompt $\pmb{p}_{ij}^e$ and $\pmb{p}_{ij}^d$, we initialize $h\times w$ to $32\times32$, and the dimension $c$ remains consistent with the dimension of the current image feature. It is worth noting that although the CSP module is more complex compared to other adaptive modules, its parameter design is relatively lightweight. Therefore, as shown in Table \ref{tab2}, its inference speed is comparable to similar architectures. For each model, we use the Adam optimizer with a learning rate of $10^{-4}$. The batch size is set to 10, and the model requires approximately one week of training on a single RTX 3090 GPU to complete training.

\noindent\textbf{3)Evaluation Metrics}

The baseline includes traditional separable encoding schemes and novel DeepJSCC models. Specifically, traditional codecs use BPG and 5G LDPC with a block length of 6144 bits. We also considered channel coding under ideal capacity conditions for AWGN channels, represented as BPG+capacity. Additionally, DeepJSCC \cite{bourtsoulatze2019deep}, ADJSCC \cite{xu2021wireless}, and WITT \cite{yang2023witt} are chosen as deep learning baselines. Among them, DeepJSCC is the original deep learning based JSCC encoding scheme, which uses convolutional neural networks as the encoding and decoding network. On the basis of DeepJSCC, ADJSCC designed an AF module using the FC layer for SNR adaptation. WITT uses transformer as the backbone of the codec and utilizes Channel Module as SNR adaptive module. Both DeepJSCC and ADJSCC have been improved with state-of-the-art modules and large dimensions.

\subsection{Result analysis}
The PSNR performances vary with SNR on three datasets under AWGN channel are illustrated in Fig. \ref{fig:PSNR} (a), (b), (c), while the results under Rayleigh fading channel are in Fig. \ref{fig:PSNR} (d), (e), (f). Notably, the proposed PJSCC outperforms the deep learning based counterparts across the entire SNR regime. Under low SNR regimes, the performances of PJSCC-U and PJSCC-A/PJSCC-R are similar. Particularly, the proposed PJSCC showcases relatively larger enhancement over the baselines especially for high resolution images under Rayleigh fading channel by an average gain of 1.97 dB.
Compared to traditional separable coding schemes, the proposed PJSCC surpassed the performance of all BPG+LDPC combinations almost among all the SNR regimes, particularly for high resolution images. The PJSCC does not suffer from the cliff effect and showcases an absolute advantage in complex environments by attaining a qualitative leap in terms of PSNR.

Similarly, the reconstruction performances in terms of perceptual metric LPIPS are demonstrated in Fig. \ref{fig:LPIPS}. The proposed PJSCC surpasses the baselines under almost all conditions in terms of  visual reconstruction quality, which is crucial in semantic communications. Fig. \ref{fig:Visualization} showcases reconstruction results from Kodak dataset of different coding schemes under the same CBR and SNR condition. The proposed PJSCC holds clearer details and achieves both higher qualitative and quantitative performance.

\begin{table}[t]
\caption{The inference speed and number of computations of different coding schemes on different resolution images.}
\begin{center}
\begin{tabular}{c c c c c}
\cline{1-5} 
\textbf{}&\multicolumn{2}{c}{\textbf{High Resolution Image}}& \multicolumn{2}{c}{\textbf{Low Resolution Image}}\\
Model & Inference Time&FLOPs& Inference Time& FLOPs \\
\hline
PJSCC&52.3ms&39.5G&62.4us&767M\\
WITT\cite{yang2023witt}&45.8ms&33.6G&74.1us&803M\\
ADJSCC\cite{xu2021wireless}&139ms&510G&111us&11.5G\\
DeepJSCC\cite{bourtsoulatze2019deep}&121ms&509G&89.1us&11.4G\\
\hline
\end{tabular}
\label{tab2}
\end{center}
\end{table}

\begin{table}[t]
\caption{Comparison of storage usage of high resolution image models under different amount of channel distribution. }
\begin{center}
\begin{tabularx}{0.75\linewidth}{>{\centering\arraybackslash}X >{\centering\arraybackslash}X >{\centering\arraybackslash}X}
\cline{1-3} 
\hline
{}& {\textbf{One Channel}}&{\textbf{Two Channel}}\\
{Model}& {\textbf{Distribution}}&{\textbf{Distribution}}\\
\hline
PJSCC&178M&216M\\
WITT\cite{yang2023witt}&120M&240M\\
ADJSCC\cite{xu2021wireless}&131M&262M\\
DeepJSCC\cite{bourtsoulatze2019deep}&1469M&2938M\\
\hline
\end{tabularx}
\label{tab3}
\end{center}
\end{table}

Table \ref{tab2} showcases the inference time and computational demands in terms of Floating Point Operations (FLOPs) of multiple DeepJSCC schemes on different resolution images. All experiments are conducted on a single RTX3090 GPU. Notably, when transmitting low-resolution images like CIFAR-10, PJSCC demonstrates superior computational efficiency, with inference times lower than all baseline methods. The PJSCC showcases a slight decrease in inference speed for larger images in Kodak/CLIC since more CSP modules are used. Note that PJSCC and WITT \cite{yang2023witt} hold comparable FLOPs, however merely one PJSCC model is suitable for various physical channel conditions and multiple WITT models are required to be trained. Additionally, the ADJSCC \cite{xu2021wireless} and DeepJSCC \cite{bourtsoulatze2019deep} cost more than ten times of FLOPs, which is will bring huge training costs.

To further demonstrate the advantages of our model in practical applications and deployment, we compared the storage costs of models used for high-resolution images, as shown in Table \ref{tab3}. The proposed model trained for single channel distribution is slightly larger than baselines. Nonetheless, the universal model PJSCC-U trained for both AWGN and Rayleigh fading channel is smaller than baselines. This is because only more prompts are required for the training of PJSCC-U, while the baselines require the preparation of two separate models.  Hence, the proposed PJSCC demonstrates good scalability and memory advantages. In practical use, this also reduces the cost associated with model switching.


\begin{table}[h]
\caption{
The PSNR performances of ablation models under AWGN channel and CBR=1/3 on the CIFAR-10 dataset. (Unit: dB)}
\begin{center}
\begin{tabular}{c c c c c c}
\hline
Method & \textbf{SNR=1}& \textbf{SNR=4}& \textbf{SNR=7}& \textbf{SNR=10} & \textbf{SNR=13}\\
\hline
w CSP&\pmb{30.30}&\pmb{33.28}&\pmb{35.92}&\pmb{38.17}&\pmb{39.97}\\
w AF&29.64&32.64&34.84&36.83&38.58\\
w/o CSP&29.56&32.50&34.68&36.65&38.40\\
\hline
\end{tabular}
\label{tab1}
\end{center}
\end{table}

We present the results of ablation study in Table \ref{tab1} to evaluate the learnable physical channel prompt. We test the effect of CSP module by removing it out under multiple SNR conditions. It can be observed that the PSNR performance will be degraded without CSP.  We also compare with the attention feature (AF) module proposed in \cite{xu2021wireless} that directly introduces SNR conditions into the communication system. It can be observed that the SNR adaptive capability of CSP module outperforms AF module by at least 1 dB. Therefore, we can conclude that the CSP module is flexible and has superior generalization ability.

\section{Conclusion}
In this paper, we have presented a novel approach, PJSCC, to enhance the capacity of physical channel adaptability in wireless image semantic transmission. PJSCC leverages DeepJSCC principles to enable semantic communications across various SNR and multiple channel distributions without necessitating retraining. 
We have proposed the innovative CSP module that effectively learns a set of prompts to incorporate channel conditions implicitly into the transmission network.  Comprehensive experiments have demonstrated the robustness of PJSCC in adapting to diverse SNR conditions, encompassing both AWGN and Rayleigh channels, while consistently achieving optimal reconstruction performance across multiple metrics. Notably, PJSCC exhibits memory efficiency and facile deployment feasibility in practical scenarios. Future work will encompass more real-world channel scenarios and further explore the potential of semantic communications.

\section*{Acknowledgment}

This work is supported in part by the National Natural Science Foundation of China (NSFC) under Grant 62101307, U23B2052, 62321001, 92267202, 62341104 and in part by the Fundamental Research Funds for the Central Universities 2023RC78.

\bibliographystyle{IEEEtran}
\bibliography{globalcom}

\end{document}